\documentclass[a4paper,floatfix,superscriptaddress,twocolumn,aps,prb,showpacs,longbibliography]{revtex4-2}
\usepackage{graphicx}
\usepackage{dcolumn}
\usepackage{bm}
\usepackage{wrapfig}
\usepackage{soul}
\usepackage[dvipsnames]{xcolor}
\usepackage{dcolumn}
\usepackage{bm}
\usepackage[colorlinks=true,citecolor=blue,linkcolor=blue,pdfhighlight =/O]{hyperref}
\usepackage{amsmath}
\usepackage{cancel}
\usepackage{dsfont}
\usepackage{mathtools}
\usepackage[dvipsnames]{xcolor}
\usepackage{soul}
\usepackage[normalem]{ulem}
\usepackage{bm}
\usepackage{amsmath,amssymb,amsfonts,latexsym,fancyhdr,graphicx}
\usepackage{simplewick}
\usepackage[english]{babel}
\usepackage{graphicx}
\usepackage{subfigure}
\usepackage{xcolor}
\usepackage{color}
\usepackage{verbatim}
\usepackage{units}

\begin{document}

\title{Interplay of Andreev reflection and Coulomb blockade in hybrid superconducting single electron transistors}
\author{Laura Sobral Rey}%
\affiliation{Physics Department, University of Konstanz, 78457 Konstanz, Germany}
\author{David Christian Ohnmacht}%
 \affiliation{Physics Department, University of Konstanz, 78457 Konstanz, Germany}
 \author{Clemens B. Winkelmann}%

\affiliation{\mbox{Univ.} Grenoble Alpes, CNRS, Grenoble INP, Institut N\' eel, 38000 Grenoble, France}
\author{Jens Siewert}%
\affiliation{University of the Basque Country UPV/EHU and EHU Quantum Center, 48080 Bilbao, Spain}
\affiliation{Ikerbasque, Basque Foundation for Science, 48009 Bilbao, Spain}
  \author{Wolfgang Belzig}%
 \affiliation{Physics Department, University of Konstanz, 78457 Konstanz, Germany}
\author{Elke Scheer}
 \email{elke.scheer@uni-konstanz.de}
 \affiliation{Physics Department, University of Konstanz, 78457 Konstanz, Germany}
 
\date{\today}

\begin{abstract}

We study the interplay between Coulomb blockade and superconductivity in a tunable superconductor-superconductor-normal metal single-electron transistor. The device is realized by connecting the superconducting island via an oxide barrier to the normal metal lead and with a break junction to the superconducting lead. The latter enables Cooper pair transport and (multiple) Andreev reflection. We show that those processes are relevant also far above the superconducting gap and that signatures of Coulomb blockade may reoccur at high bias while they are absent for small bias in the strong-coupling regime. 
Our experimental findings agree with simulations using a master equation approach in combination with the full counting statistics of multiple Andreev reflection.

\end{abstract}

\maketitle

Coulomb blockade (CB) is an archetypical manifestation of charge quantization (CQ), which occurs in the electronic transport across a small metallic island \cite{SingleCharge92}. CB can be suppressed by both classical and quantum fluctuations of the charge \cite{QuantumTransport09}. Classical charge fluctuations originate from thermal activation over an energy barrier, provided by the charging energy $E_{\rm C} = e^2/2C$, with $C$ the total capacitance and $e$ the elementary charge. The quantum fluctuations 
can be described using the Landauer-Büttiker 
picture using individual conductance channels $i = 1\dots N$, with transmissions $0\leq\tau_i\leq1$, connecting the island to the leads \cite{QuantumTransport09}. In the case of normal metallic leads, 
the magnitude of the conductance oscillations in a Coulomb-blockaded island in the single electron transistor (SET) geometry \cite{likharev87,FultonDolan87} was predicted to scale like $\prod_i \sqrt{1-\tau_i} \exp(-k_\text{B}T/E_\text{C})$ \cite{NazarovCBnoTB99}, which was shown to hold experimentally with great accuracy \cite{Jezouin16}. In the presence of superconducting (S) contacts this picture can be expected to change radically due to the different nature of the charge carriers.

For mesoscopic transport processes involving S leads, the energy-dependent quasiparticle (QP) spectrum has a gap $\Delta$, which strongly inhibits tunneling at low energies and small 
$\tau$ \cite{tinkham}. At larger $\tau$, multiparticle superconductive transport (MST) \cite{tinkham, btk82, Courtois99} comes into play. These processes can be either coherent, in the form of Josephson transport of Cooper pairs (CPs), in the absence of a voltage drop, or they may be dissipative, as, e.g., multiple Andreev reflection (MAR) of order $m$, which sets in above a bias voltage threshold $eV = 2\Delta/m$. Importantly, while MAR is itself disspative and therefore incoherent, 
the transfer of $m$ charges involved in such a MAR is a 
coherent 
tunneling process. The theoretical description and to-date experimental realizations of MAR were based on superconducting contacts described by a well-defined quantum phase, i.e., large particle number fluctuations.

From the above discussion, it is clear that MST 
and CQ are antagonistic processes. On the one hand, CQ effects should deeply modify the energy thresholds for a $m$-particle tunneling process On the other hand, MST could overthrow the CQ conditions known from the case of normal leads.

\begin{figure}
    \centering
    \includegraphics[width=0.35\textwidth]{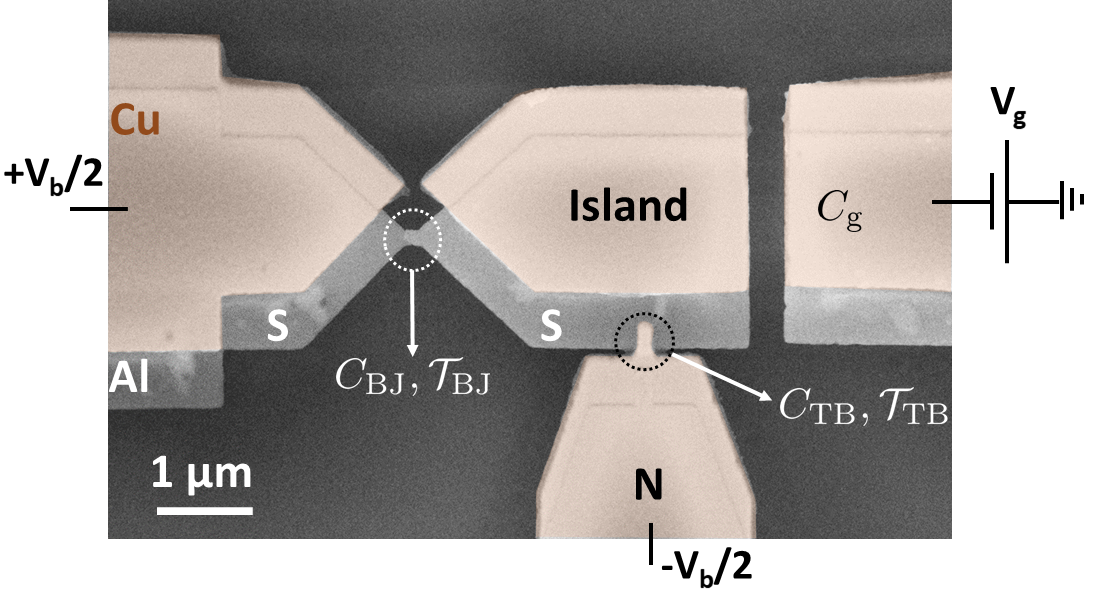}
    \caption{Scanning electron microscope picture of a single electron transistor (SET) with a break junction (BJ) as one of the tunnel junctions in series with a superconductor-insulator-normal metal tunnel junction (TB), and the area in the middle (island) being capacitively coupled to a gate electrode.
    }
    \label{figExpSample}
\end{figure}

In this Letter, we address 
the delicate interplay of CQ and MST. To this end we build a superconductor-superconductor-normal metal (SSN)-SET, hosting 
a superconducting break junction (BJ), with a small set of mechanically tunable transmission channels. The BJ is connected in series with a superconductor-normal metal (SN) tunnel barrier (TB), 
 providing thus the conditions for CQ in the S island formed between the BJ and the TB, see Fig.~\ref{figExpSample}. The device is completed by a capacitively coupled gate electrode \cite{likharev87,FultonDolan87}. In the limit of a small total transmission $\mathcal{T}_\text{BJ} = \Sigma_i \tau_{i,\text{BJ}}\ll 1$ of the  BJ, we study how the extra energy cost arising from CB 
 affects the conductance thresholds for 
the different MAR processes. 
 Furthermore, in the limit of a larger $\mathcal{T}_\text{BJ}\sim 1$, we observe that CQ enters in competition with the Josephson effect across the 
BJ, leading to the complete suppression of CB at low bias, whereas it is entirely restored 
beyond the Josephson critical current. Our data are well described by a 
master equation approach including CB as well as MST.

The transport in the SET is characterized by current cycles 
composed of charge transfer processes through the individual junctions comprising different types and different number of charged particles \cite{fitzgerald98}. The $e$-cycle is a process that charges the island with one QP through one junction, and discharges it through the other junction with another QP,  leaving the island in its initial charge state. AR and MAR cycles charge the island with $m$ electron charges through one junction, and subsequent single-QP 
 or (M)AR processes discharge the island through the other junction. The Josephson quasiparticle (JQP) cycle \cite{Averin86,NakamuraTsai96,pohlen00,manninen97} %
is a coherent process in which a CP tunnels through one junction, and simultaneously one QP tunnels off through the other junction. The initial charge state is reached by a subsequent single-QP process. 
The JQP cycle is resonant, i.e., it appears as a peak in the current-bias voltage ($I-V_\text{b}$) characteristics.

Studies on all-superconducting (SSS)-SETs in different realizations were conducted in the weak-coupling regime in which both junctions were realized as tunnel contacts and $e$, JQP and processes comprising 3 QPs were found \cite{fitzgerald98}. Another realization using a tunable junction showed a multitude of current cycles which were difficult to identify \cite{lorenz18}. Hence the SSN-SET studied here is chosen to enable MST predominantly through just one of the junctions (BJ), thereby reducing the number of processes.
While some theoretical studies on the SSN-SET exist \cite{Schon95, SiewertSSN}, to the best of our knowledge, it has never been investigated experimentally before.

For forming the tunable junction we use the thin-film BJ technique \cite{vanruitenbeek96}. The BJ consists of an 
Al bridge suspended above a flexible substrate: by bending the substrate it is possible to 
adjust contacts with arbitrary transmission $\mathcal{T}_\text{BJ}$ 
from the tunnel regime $\mathcal{T}_\text{BJ}\ll 1$ 
to atomic-size contacts with $\mathcal{T}_\text{BJ}\gtrsim 1$ \cite{ElkeAlu97, agrait03}. 
The TB with fixed resistance $R_\text{TB}$ is formed by an Al$_x$O$_y$ layer between the S (Al) and the N (Cu) material 
and hosts many channels in parallel with very small transmission $\tau_i \ll 1$ \cite{QuantumTransport09}. The electrostatic energy necessary to charge the island is 
 $   \mathcal{E}(n_{\rm g},n) = E_\text{C}(n-n_\text{g})^2$
where $n$ is the excess charge (in units of $e$) of the island, $n_\text{g}$ is the equivalent charge induced by the gate voltage, and $C = C_\text{BJ}+C_\text{TB}+C_\text{g}$ the total capacitance of the island. Furthermore we define the capacitance divisors $\kappa_i = (C_i+C_\text{g}/2)/C$. 
In this Letter, we discuss data taken on one sample.
The sample preparation and parameters as well as data on three other samples with similar properties are given in the Supplemental Material (SM) \cite{SM}. 

\begin{figure}[thb]
    \centering
    \includegraphics[width=0.45\textwidth]{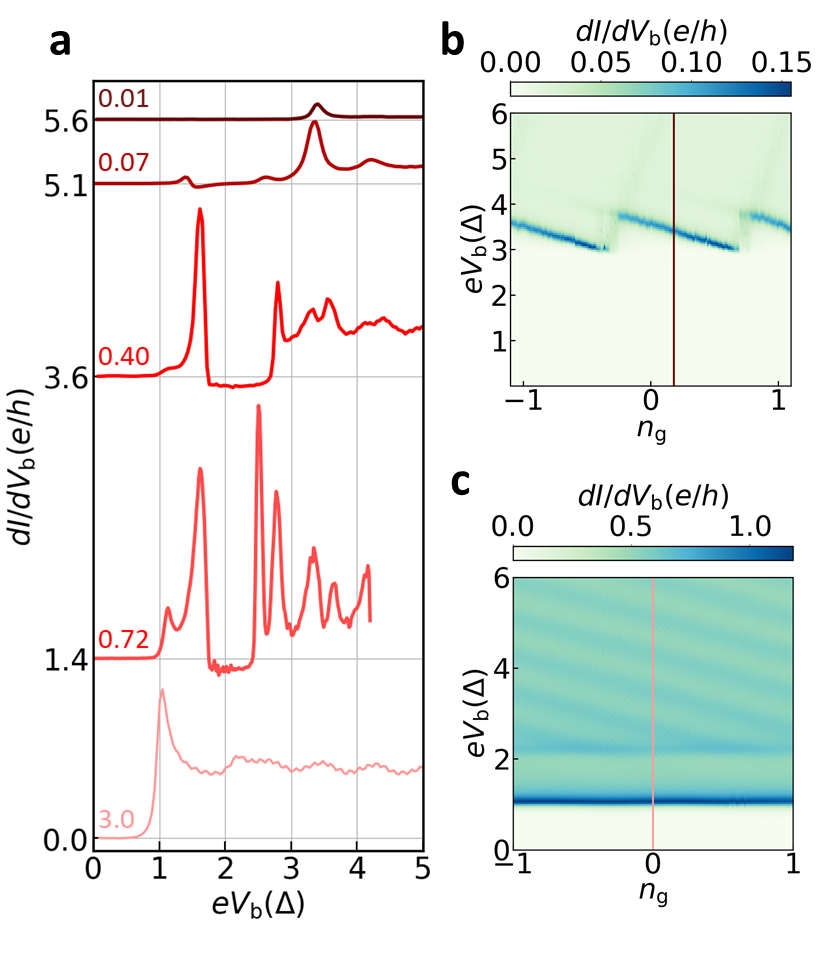}
        \caption{Evolution of the transport processes in the SSN-SET for different $\mathcal{T}_\text{BJ}$: (a) $dI/dV_\text{b}-eV_\text{b}$ cuts for varying $\mathcal{T}_\text{BJ}$  as indicated in the legend. $n_\text{g}-eV_\text{b}-dI/dV_\text{b}$ maps for $\mathcal{T}_\text{BJ} = 0.01$ (b) and $\mathcal{T}_\text{BJ} = 3$ (c). 
        }   
    \label{figComparison}
\end{figure}
\begin{figure}
    \centering
    \includegraphics[width=0.45\textwidth]{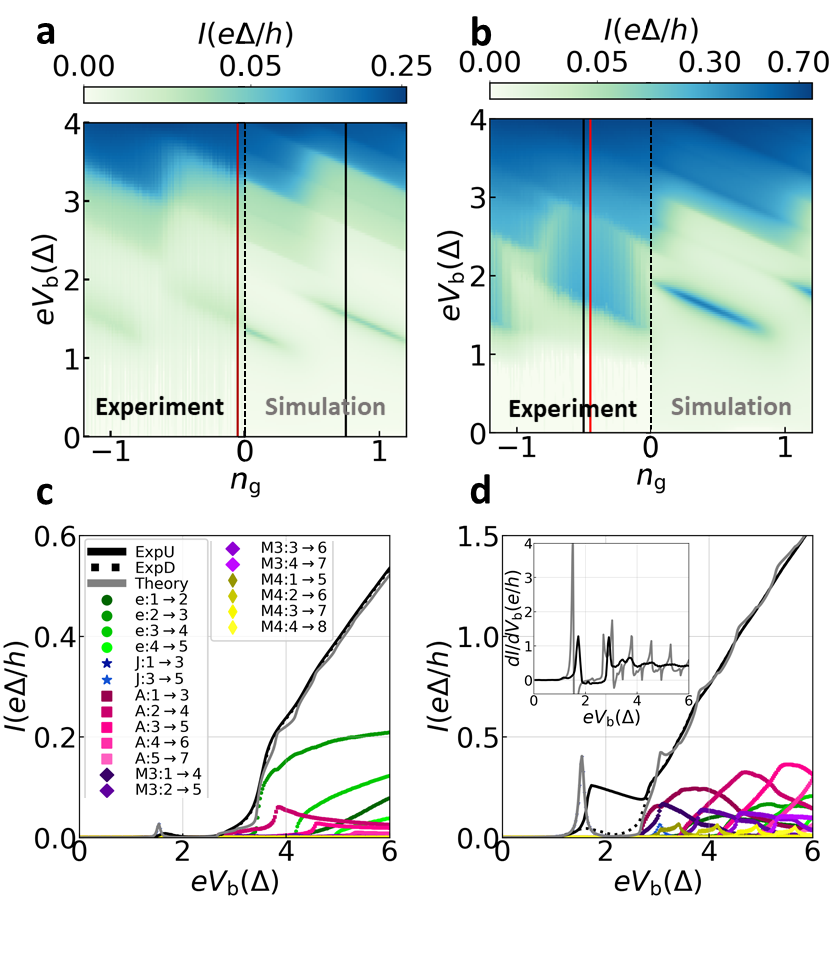}
    \caption{
    Theoretical and experimental $n_\text{g}-eV_\text{b}-I$ maps for a contact with $\mathcal{T}_\text{BJ} =$ 0.07 (a) and $\mathcal{T}_\text{BJ}$ = 0.4 (b). Red lines in  (a) and (b): positions of cuts of Fig.~\ref{figComparison}. (c,d): correspondent $I-eV_\text{b}$ cuts at constant $n_\text{g}$
    as marked in (a) and (b) with solid black lines.
    Black/gray: experiment (solid/dashed line: increasing/decreasing $|V_\text{b}|$) /simulation. Scatter plots: $P_{n} \Gamma_{n\rightarrow n+n'}$ contributions (with amplitude $> 0.05 e\Delta/h$): $e$ ($e$: $n\rightarrow n+1$, green), AR (A: $n\rightarrow n+2$, pink), $m = 3$ MAR (M3: $n\rightarrow n+3$, purple), $m = 4$ MAR (M4: $n\rightarrow n+4$, yellow). 
   . 
    Insert in (d):  experimental (black) and simulated (gray) $dI/dV_\text{b}$. }
    \label{figEvolution}
\end{figure}

\begin{figure}
    \centering
    \includegraphics[width=0.45\textwidth]{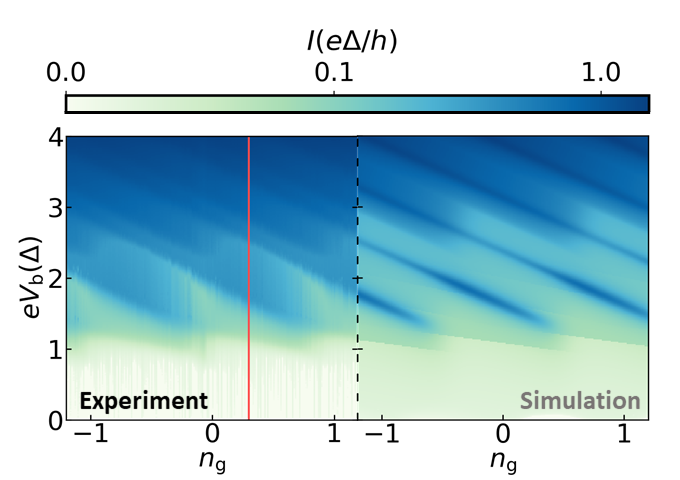}
    \caption{Experimental and simulated maps for a contact with $\mathcal{T}_\text{BJ}$ = 0.72. The red line marks the position at which the line cut shown in Fig.~2(a) has been taken.
    }
    \label{figHighOrderProcess}
\end{figure}

  
We start the discussion by comparing different coupling strengths for the BJ, from weak to strong. 
Fig.~\ref{figComparison}(a) shows $dI/dV_\text{b}-eV_\text{b}$ plots taken at fixed $n_\text{g}$ for different $\mathcal{T}_\text{BJ}$ (from top to bottom) of 0.01, 0.07, 0.4, 0.72 and 3. From these individual conductance curves taken at different $n_\text{g}$ we construct conductance vs. gate voltage and bias voltage maps in reduced units as  $n_\text{g}-V_\text{b}-dI/dV_\text{b}$ color plots. Two extreme examples, for $\mathcal{T}_\text{BJ} = 0.01$  and 3, are given in Fig.~\ref{figComparison}(b,c), respectively. For the other transmissions, we plot the $n_\text{g}-eV_\text{b}-I$ in Fig.~\ref{figEvolution}(a,b) ($\mathcal{T}_\text{BJ} = 0.07$ and 0.4, respectively) and Fig.~\ref{figHighOrderProcess} for $\mathcal{T}_\text{BJ} = 0.72$. Further examples are shown in the SM \cite{SM}.  The maps show that different current contributions are modulated by the gate voltage with $e$-periodicity with asymmetric slopes indicating different capacitances of the junctions. The positive/negative slope mark the onset of charge transport across the TB/BJ, respectively.  The visibility of the transport across the BJ is much higher than the one across the TB as a consequence of the different nature of the junctions. The vertical red lines mark the positions at which the data in Fig.~\ref{figComparison}(a) have been taken. They correspond to positions in the middle of the descending slope of a CB diamond \footnote{We show here data for positive $V_\text{b}$, but note that all maps were symmetric with respect to the bias polarity when choosing the adequate $n_{\rm g}$. For intermediate $\mathcal{T}_\text{BJ} \sim 0.5$ the peak slightly above $\Delta$ is hysteretic with respect to the bias sweep direction.}. 
For more details see the SM \cite{SM}.

For the lowest $\mathcal{T}_\text{BJ} = 0.01$ in Fig.~\ref{figComparison}(a), 
we observe only one clearly visible peak at $3\Delta$, marking the $e$-process. 
For $\mathcal{T}_\text{BJ} = 0.07$, the peak corresponding to the $e$-process is bigger in amplitude, and two additional peaks appear between $\Delta$ and $3 \Delta$. The first, that --  depending on $n_{\rm g}$ -- is located between $\Delta+E_\text{C}$ and $\Delta+3E_\text{C}$ ($\sim$1.4$\Delta$ and 2.2$\Delta$), corresponds to a JQP cycle \cite{lorenz18,fitzgerald98, nakamura95,pohlen00}, while we assign the second one (with a lower threshold of $2\Delta +E_\text{C}$) to a cycle involving AR across the BJ. Finally, the peak at around $4.5 \Delta$ signals again the next step in the Coulomb staircase.  
For $\mathcal{T}_\text{BJ} = 0.4$ the amplitude of the JQP and the AR cycle increase further. At the position of the $e$-process a double peak occurs that we identify as an AR process followed by a MAR process with $m = 3$, 
indicating that both AR and MAR are relevant processes in this coupling regime. 
For $\mathcal{T}_\text{BJ} = 0.72$ the amplitude of the MAR and JQP cycles has further increased  such that the JQP of the adjacent diamond, i.e., signaling the next charge state, gives rise to the highest peak at around 2.4$\Delta$.
There is no particular peak starting at 3$\Delta$, meaning that cycles including MST are more relevant than $e$ transport. Besides, there is a small peak slightly above $\Delta$ that can be explained by considering cotunneling of several QPs, see below.
For large $\mathcal{T}_\text{BJ}$
large Josephson currents 
can flow without a voltage drop. In this situation
the voltage drops completely across the SN-TB, and the island’s chemical potential remains pinned to that of the S contact. This is visible in the lowest trace in Fig.~\ref{figComparison}(a) as well as in Fig.~\ref{figComparison}(c), where
above $\Delta/e$ no gate dependence is observed, indicating a suppression of CQ. 
Upon further increasing $|eV_{\rm b}|$ to about $2.2\Delta$, 
gate-dependent conductance features with the same slope as observed for small $\mathcal{T}_\text{BJ}$ reappear. At this threshold, the current amounts to 7.0 nA, an order of magnitude below the 
estimation $I_\text{c} = \frac{\pi}{2} G_0 \tau \Delta\approx 69$ nA \cite{AmbegaokarBaratoff} of the maximally possible supercurrent, in agreement with reports from earlier experimental \cite{joyez1994,cleuziou2006} and theoretical works \cite{vandenbrink1991} in SSS-SETs. As the dissipation-free state of the SS junction is lost, the island’s chemical potential is no longer pinned via the Josephson effect to that of the S lead. Consequently, the SS junction switches from a vanishing to a high impedance.
This restores CB effects 
despite of the large $\mathcal{T}_\text{BJ}\approx3$.

To verify this 
assignment of the various contributions, we calculate the current through the device with the master equation approach (Eq.~\ref{MasterEquation}) \cite{SingleCharge92}. We take into account the JQP rate $\Gamma_\text{JQP} (\delta E)$ from Averin and Aleshkin \cite{averinAleshkin} and rates derived from the full counting statistics for (M)AR processes $\Gamma_\text{MAR} (\delta E)$ \cite{CuevasFCS03} in the BJ. For the TB, we solely use single-QP transport using the classical rate expression for a SN junction \cite{tinkham} \footnote{For describing the cotunneling processes at intermediate coupling we will also take a small AR contribution through the TB junction into account.}. The electrostatic energy difference can be expressed as \cite{SingleCharge92}:

\begin{equation}
\delta E = \mathcal{E}(n_\text{g},n)-\mathcal{E}(n_\text{g},n+n')+ n' \kappa_i eV_\text{b}
\end{equation} 
where $n'$ is the number of transferred charges. 
We assume that there is no accumulation of electrons in the island or relevant environment and neglect effects of the environment \cite{pekola10}, so $dP/dt = 0$ and the current through the two junctions is equal. Considering the normalization condition $\sum_n P_n = 1$ we solve the 
stationary master equation Eq.~\eqref{MasterEquation} and obtain:

\begin{equation}
\sum_{n' \neq 0}P_{n+n'} \Gamma_{n+n'\rightarrow n}\ -\ P_{n} \sum_{n' \neq 0} \Gamma_{n\rightarrow n+n'}\ =\ 0
\ \ .
\label{MasterEquation}
\end{equation}

From these, 
we calculate the total current through the SSN-SET by considering the charge transferred, e.g., in the BJ,
\begin{equation}
    \begin{split}
I\ =\ (-e) \sum_{n; n'\neq 0} 
  \tilde{n}\ P_{n} \ \Gamma_{\mathrm{BJ},n\rightarrow n+n'}
    \ \ .
\label{EquationCurrent}
\end{split}
\end{equation}
\noindent Here, $n'$ can also be negative, and 
$\tilde{n}$ is the charge transferred in the BJ
(which may be different from $n'$).

Figs.~\ref{figEvolution}(a) and (c) are the $n_\text{g}-eV_\text{b}-I$ map and a $I-eV_\text{b}$ cut at $n_\text{g}$ = 0.75 as indicated in the map, for a configuration with a $\mathcal{T}_\text{BJ} = 0.07$ for the BJ. 
In Fig.~\ref{figEvolution}(c), the black lines correspond to experimental curves (solid/dotted for increasing/decreasing voltage) while the gray lines correspond to the simulation. The colored symbols depict the individual contributions $P_{n} \Gamma_{n\rightarrow n+n'}$ that sum up to the total current
The resonant peak measured in the experiment follows the JQP contribution from the simulation, and the small feature starting at 2.4$\Delta$ is due to a cycle with AR 
\cite{fitzgerald98, maisi2011}. 
Fig.~\ref{figEvolution}(b,d) show the comparison of experiment and theory for a higher transmission $\mathcal{T}_\text{BJ} = 0.4$. As Fig.~\ref{figEvolution}(b) shows, the AR, MAR and JQP cycles become more pronounced and the $e$-cycle starting at $3 \Delta$ is almost invisible. The large blue area in the experimental map above the JQP line (at $\sim 1.5\Delta$) is a result of the resonant character of the JQP and is absent for the downsweep (see SM). The decomposition into individual cycles in Fig.~\ref{figEvolution}(d) 
confirms that  AR and MAR with $m = 3$ (pink and purple symbols) are bigger in amplitude than the $e$ transport (green symbols). In single SS junctions $e$ and AR persist to very high bias, while MAR fades out above $2\Delta$ \cite{CuevasFCS03}. In contrast, here, also MAR contributes to the current far outside the gap region. This finding shows that MAR processes 
become more robust by the interplay with CB. 
The $dI/dV_\text{b}$ in the inset shows that the multipeak characteristic of each charge state found in the simulation, indicative for the presence of the MAR processes, is also observed in the experimental curve
\footnote{We note that Al single-atom contacts with $\mathcal{T}\gtrsim 0.8$ are expected to accommodate more than one channel \cite{ElkeAlu97}. For this contact with $\mathcal{T} = 0.4$, it was sufficient to consider one single channel with $\tau = 0.4$. Considering more than one channel would result in a smaller current, and hence, would not improve the agreement between experiment and simulation giving a worse agreement.}. 
For even higher $\mathcal{T}_\text{BJ}$ (Fig.~\ref{figHighOrderProcess}) the amplitudes of the  processes setting in below $3\Delta$ increase, and besides a new line appears at $|eV_\text{b}| = \Delta$. In the $n_\text{g}-eV_\text{b}$ map the slope of this cycle is around 3 times smaller than the one of the other cycles. 

Here, the lowest-order cycle with the observed slope consists of a coherent tunneling process that simultaneously transports three charges through the BJ and two through the TB, hence implies AR in the TB. This process has the following electrostatic energy balance   
 $   \delta E = \mathcal{E}(n_\text{g},n)-\mathcal{E}(n_\text{g},n+1)+(3\kappa_\text{BJ}+2\kappa_\text{TB}) eV_\text{b}$.
This equation implies that the island charge changes by $e$ and 
shows that the slope in the $n_\text{g}-eV_\text{b}$ map is a function of the capacitance ratio and the number of charge carriers involved ~\cite{Joyez1995,SS1996}. For the sample studied here it turns out to be 1/2.88, hence close to 1/3.   
The expression for such a rate has not been worked out theoretically. Therefore, we tentatively assume the rate as a step function starting at $\delta E = 2\Delta$ with an amplitude as free parameter.
In Fig.~\ref{figHighOrderProcess} we show the comparison between experiment and theory for a contact with $\mathcal{T}_\text{BJ}$ = 0.72 that shows this feature starting at $|eV_\text{b}| = \Delta$ and with a slope  
in agreement with the experimental result. 
The contributions at higher bias, namely the JQP and MAR cycles, appear in the simulation at around the same position as in the experimental map and they follow the same slopes. The amplitude of both processes overcomes the $e$ transport for high voltages and extends to very high $|V_\text{b}|$. The apparent double-line pattern above $2.5\Delta$ is the result of JQP and MAR cycles from different initial charge states of the island. We note that we also included the three-QP cotunneling process  (AR in the BJ and $e$ in the TB) in the simulation. This contribution sets in at $eV_\text{b}\sim1.5\Delta $, has a slope of $\sim 1/2$  and its onset is therefore hidden by the JQP contribution. However, it contributes to the overall current level, improving the agreement between theory and experiment. Simulation results without these contributions are shown in the SM \cite{SM}. 

In conclusion, we presented transport properties of an SSN-SET 
with a tunable junction allowing to vary the relative strength of $e$, JQP, AR and MAR processes.  With this approach we show the mutual influence between MST processes and CB. While in single junctions MAR processes are not relevant outside the gap region and only $e$ and AR processes make up the excess current, the combination with CB enhances the contributions of MAR at high bias.  We show that the description based on charge states of the island allows to quantitatively reproduce the experimental data up to relatively strong coupling $\mathcal{T}_\text{BJ}\simeq 1$ when taking all four kinds of cycles into account. In particular MAR cycles show up in three different ways in the SET: 
MAR of order $m$ charges the island with $me$ and its amplitude can be higher than the one for $e$ transport.  AR and MAR are the dominating transport process up to very high voltages. For higher $\mathcal{T}_\text{BJ}$, cotunneling events become relevant and it is possible to identify, qualitatively, a  coherent multiparticle cotunneling process across  both junctions inside the gap region $|eV_\text{b}|\leq 3\Delta$, which implies MAR with $m=3$ in the BJ and AR in the TB. The third and most striking signature of MAR is the reappearance of charging effects at high bias in the strong coupling regime ($\mathcal{T}_\text{BJ}>1$) when at least one channel with a transmission $\tau = 1$ exists in the BJ) while at low bias all signature of CQ has already disappeared. 
Our findings are important for designing and understanding nanoscale superconducting quantum devices,  such as  superconducting qubits \cite{TransmonReview}, with arbitrary intermediate transmissions to use the subtle interplay between the charge and phase degrees of freedom as a resource for optimizing their performance.\\

Acknowledgements: We thank J. C. Cuevas  for fruitful discussion and S. Sprenger and T. Lorenz-Sprenger for experimental support. We gratefully acknowledge funding by the European Commission through the Marie Skłodowska-Curie grant agreement no. 766025 (QuESTech)  and the Deutsche Forschungsgemeinschaft (DFG; German Research Foundation) by SFB 1432 (Project No. 425217212).

%



\end{document}